\documentclass[twocolumn]{aastex61}

\newcommand{\integral} {{\em INTEGRAL}}

\newcommand{\swift}{{\em Swift}}
\newcommand{\chandra}{{\em Chandra}}
\newcommand{\hst}{{\em HST}}

\newcommand{\swiftxrt}{{\em Swift/XRT}}

\newcommand{\IGR}{{IGR~J16597$-$3704}}
\newcommand{\chandralong}{{\em Chandra X-ray Observatory}}

\shorttitle{\IGR{}}
\shortauthors{Tetarenko, A.J. et al.}

\begin{document}


\title{A radio frequency study of the accreting millisecond X-ray\\pulsar, \IGR{}, in the globular cluster NGC 6256}


\author[0000-0003-3906-4354]{A.J. Tetarenko}
\affiliation{Department of Physics, University of Alberta, CCIS 4-181, Edmonton, AB T6G 2E1, Canada}

\author[0000-0003-2506-6041]{A. Bahramian}
\affiliation{Department of Physics and Astronomy, Michigan State University, East Lansing, MI, USA}

\author[0000-0002-3516-2152]{R. Wijnands}
\affiliation{Anton Pannekoek Institute for Astronomy, University of Amsterdam, Science Park 904, 1098 XH Amsterdam, the Netherlands}

\author[0000-0003-3944-6109]{C.O. Heinke}
\affiliation{Department of Physics, University of Alberta, CCIS 4-181, Edmonton, AB T6G 2E1, Canada}

\author[0000-0003-0976-4755]{T.J. Maccarone}
\affiliation{Department of Physics and Astronomy, Texas Tech University, Box 41051, Lubbock, TX 79409-1051, USA}

\author[0000-0003-3124-2814]{J.C.A. Miller-Jones}
\affiliation{International Centre for Radio Astronomy Research, Curtin University, G.P.O. Box U1987, Perth, WA, 6845, Australia}

\author[0000-0002-1468-9668]{J. Strader}
\affiliation{Department of Physics and Astronomy, Michigan State University, East Lansing, MI, USA}

\author[0000-0002-8400-3705]{L. Chomiuk}
\affiliation{Department of Physics and Astronomy, Michigan State University, East Lansing, MI, USA}

\author{N. Degenaar}
\affiliation{Anton Pannekoek Institute for Astronomy, University of Amsterdam, Science Park 904, 1098 XH Amsterdam, the Netherlands}

\author[0000-0001-6682-916X]{G.R. Sivakoff}
\affiliation{Department of Physics, University of Alberta, CCIS 4-181, Edmonton, AB T6G 2E1, Canada}

\author[0000-0002-3422-0074]{D. Altamirano}
\affiliation{Physics and Astronomy, University of Southampton, Southampton SO17 1BJ, UK}

\author[0000-0001-9434-3837]{A. T. Deller}
\affiliation{Centre for Astrophysics and Supercomputing, Swinburne University of Technology, Hawthorn, VIC 3122, Australia}

\author[0000-0002-6745-4790]{J.A. Kennea}
\affiliation{Department of Astronomy and Astrophysics, Pennsylvania State University, 525 Davey Lab, University Park, PA 16802, USA}

\author{K.L. Li}
\affiliation{Department of Physics and Astronomy, Michigan State University, East Lansing, MI, USA}

\author[0000-0002-7092-0326]{R.M. Plotkin}
\affiliation{International Centre for Radio Astronomy Research, Curtin University, G.P.O. Box U1987, Perth, WA, 6845, Australia}

\author[0000-0002-7930-2276]{T.D. Russell}
\affiliation{Anton Pannekoek Institute for Astronomy, University of Amsterdam, Science Park 904, 1098 XH Amsterdam, the Netherlands}

\author[0000-0002-8808-520X]{A.W. Shaw}
\affiliation{Department of Physics, University of Alberta, CCIS 4-181, Edmonton, AB T6G 2E1, Canada}

\correspondingauthor{A.J. Tetarenko}
\email{tetarenk@ualberta.ca}


\begin{abstract}
We present Karl G.~Jansky Very Large Array radio frequency observations of the new accreting millisecond X-ray pulsar (AMXP), \IGR{}, located in the globular cluster NGC 6256. With these data, we detect a radio counterpart to \IGR{}, and determine an improved source position. Pairing our radio observations with quasi-simultaneous \swiftxrt{} X-ray observations, we place \IGR{} on the radio -- X-ray luminosity plane, where we find that \IGR{} is one of the more radio-quiet neutron star low-mass X-ray binaries known to date. 
We discuss the mechanisms that may govern radio luminosity (and in turn jet production and evolution) in AMXPs. 
Further, we use our derived radio position to search for a counterpart in archival {\em Hubble Space Telescope} and \chandralong\ data, and estimate an upper limit on the X-ray luminosity of \IGR{} during quiescence.

\end{abstract}





\section{Introduction}

Relativistic jets are launched from many different types of accreting stellar-mass compact objects (black holes, neutron stars, and possibly white dwarfs; \citealt{fen06,migfen06,kording2008,coppejans2015,russ16}); however our current knowledge of the physics that gives rise to and governs jet behaviour is still somewhat limited. A crucial step towards understanding the mechanisms that drive jet behaviour is characterizing jet properties (and how these properties are coupled to the conditions in the accretion flow) in different accreting systems across the mass scale.

A key observational diagnostic for comparing jet properties between different systems is the radio -- X-ray correlation, relating radio and X-ray luminosities ($L_{\rm R}\propto L_{\rm X}^\beta$, where $\beta$ represents the disc-jet coupling index; \citealt{galfenpol03}; \citealt{corb13}).
This empirical relationship, which couples a compact, partially self-absorbed synchrotron jet (probed by radio emission) to the properties of the accretion flow (probed by X-ray emission), has been well studied in black hole X-ray binary systems (BHXBs; binary systems harbouring a black hole accreting matter from a companion star).
In particular, different BHXB systems, sampled over several orders of magnitude in X-ray luminosity, 
are known to display correlations that range from $\beta\sim0.6-1.8$ (potentially following one of two tracks in the radio X-ray plane at $L_X>10^{36}\,{\rm erg\,s}^{-1}$; \citealt{coriat11, gallo14, russelltd15}).
However, the different classes of neutron star X-ray binary systems (NSXBs; binary systems harbouring a neutron star accreting matter from a companion star) are not as well sampled (in particular due to the limited range of X-ray luminosities that have been sampled to date), and have shown  more complex behaviour in the radio -- X-ray plane, as compared to the BHXBs.

\renewcommand\tabcolsep{12pt}
\begin{deluxetable*}{ ccccc }
\tablecaption{Radio and X-ray observation details and luminosities for \IGR{}\label{table:obs_times}}
\tablehead{
\colhead{Epoch}&
\colhead{Radio Observation}&
\colhead{X-ray Observation}&
\colhead{$L_{5 {\rm GHz}}$\tablenotemark{a,b}}
&
\colhead{$L_{1\mbox{--}10 {\rm \, keV}}$\tablenotemark{a}}\\
\colhead{}&
\colhead{(mm/dd/yy, UTC)}&
\colhead{(mm/dd/yy, UTC)}&
\colhead{(${\times10^{28}}{\rm \, erg\, s}^{-1}$)}&
\colhead{(${\times10^{36}}{\rm \, erg\, s}^{-1}$)}}
\startdata
1&10/23/2017, 21:01--22:42&10/22/2017, 20:29--20:47&$0.87\pm0.22$&$2.69\pm0.11$\\
2&10/27/2017, 20:44--22:25&10/25/2017, 07:30--07:51&$1.20\pm0.21$&$2.76\pm0.08$\\
\enddata
\tablenotetext{a}{To calculate the luminosities, we use a distance to NGC 6256 of $D=9.1$ kpc \citep{val07a}. Uncertainties include measurement errors only, and are quoted at the $1\sigma$ level.}
\tablenotetext{b}{We calculate 5 GHz radio luminosities ($L_R=\nu L_\nu$) by combining the two VLA base-bands in each observation and assuming a flat spectral index to extrapolate to 5 GHz.}
\end{deluxetable*}
\renewcommand\tabcolsep{6pt}

While NSXBs are generally more radio quiet than BHXBs, different neutron star X-ray binary classes have shown varying correlation indices\footnote{We note that these correlation indices are measured over a limited range of X-ray luminosity, and \citet{corb13} found that an X-ray luminosity lever arm extending across at least 2 dex is needed to accurately measure a correlation in the radio -- X-ray plane.} and normalizations in the radio-X-ray plane \citep{mig03,migfen06,tudose09,millerj10,tetarenkoa2016,tud17}. For example, some non-pulsating neutron stars display $\beta\sim1.4$ \citep{migfen06,millerj10}, while some accreting millisecond X-ray pulsars (AMXPs; accreting neutron star binaries where X-ray pulsations at the spin period of the neutron star are observed) and three transitional millisecond X-ray pulsars (tMSPs; accreting neutron star binaries that switch between a rotation-powered pulsar state and an accretion-powered state; \citealt{arch9,papitto2013,bas14,pat14}) have been suggested to follow a shallower correlation of $\beta\sim0.7$ \citep{deller2015}. 
Further, differences are also observed between individual systems of the same class. For example, recent work has shown that not all AMXPs and non-pulsating NSXBs follow the above mentioned `standard' tracks in the radio -- X-ray plane (where some systems may display lower/higher radio luminosities; \citealt{tetarenkoa2016,tud17}).
Many different factors could play a role in causing these observed differences, such as variations in  jet power, compact object mass, spin, magnetic field, and jet launching mechanism.
To disentangle these factors, understand the reason(s) for a lack of clear correlation(s) and the wide range of radio luminosities observed in neutron star systems, constraints from a larger population of neutron star systems (especially at $L_X<10^{36}\,{\rm erg\,s}^{-1}$), are strongly needed. However, sampling neutron star systems at X-ray luminosities between $10^{34}<L_X<10^{36}\,{\rm erg\,s}^{-1}$ is observationally challenging, as neutron stars tend to evolve quickly in this luminosity range and are faint at radio frequencies.
Rapid, coordinated radio and X-ray observations of new X-ray transients discovered in our Galaxy can in principle provide these much needed constraints.

\IGR{} is a new X-ray transient discovered with the INTErnational Gamma-Ray Astrophysics Laboratory (\integral{}) on 2017 October 21 \citep{boz17a}.
Followup \swift{} X-Ray Telescope (XRT; \citealt{Burrows05}) observations \citep{boz17b} on 2017 October 22 confirmed the presence of a new bright X-ray source within the \integral{} error circle,
and placed this new transient in the globular cluster NGC 6256 ($D=9.1$ kpc; \citealt{val07a}).  
To determine the nature of IGR~J16597-3704 and localize its position, we performed Karl G. Jansky Very Large Array (VLA) radio frequency observations of IGR~J16597-3704 on 2017 October 23 and 27 \citep{teta17}. These radio observations were taken within $3$ days of \swiftxrt{} observations of the source, allowing us to also place this new source in the radio -- X-ray correlation plane. The preliminary position of \IGR{} on the radio -- X-ray plane strongly suggested that this new transient is a neutron star system. 
This classification was confirmed by \cite{sanna17}, who report the discovery of X-ray pulsations, find that \IGR{} is an ultra-compact binary ($\sim46$ minute orbital period), with a short spin period ($9.5$ ms), and suggest a high magnetic field ($9.2\times10^8<B<5.2\times10^{10}$ G).
\IGR{} was also observed with \chandra{} on 2017 October 25 \citep{chak17}.
In this paper, we report on our VLA radio and \swiftxrt{} X-ray observations, as well as our search for the optical and quiescent X-ray counterparts to \IGR{}.

\section{Observations and Data Analysis}

\subsection{VLA radio observations of \IGR{}}
\label{sec:vla}

\IGR{} was observed with the VLA (project code VLA/17B-257) over two epochs, 2017 October 23 and 27 (see Table~\ref{table:obs_times} for observation times), with 88.6 min on source at each epoch. The array was in its B configuration, with a beam size of $2.2\times0.8$ arcsec. All observations were taken using the 3-bit samplers at X-band (8--12 GHz), and were comprised of two base-bands, each with 16 spectral windows of sixty-four 2-MHz channels, providing a total bandwidth of 2.048 GHz per base-band. We carried out flagging, calibration, and imaging within the Common Astronomy Software Application package (\textsc{casa}, v5.1.1; \citealt{mcmullin2007}), using standard procedures {outlined in the \textsc{casa} Guides\footnote{\url{https://casaguides.nrao.edu}} for VLA data reduction (i.e., apriori flagging, re-quantizer gain corrections, setting the flux density scale, initial phase calibration, solving for antenna-based delays, bandpass calibration, gain calibration, scaling the amplitude gains, and final target flagging)}. When imaging we used a natural weighting scheme to maximize sensitivity and two Taylor terms (nterms=2) to account for the large fractional bandwidth. 
We used 3C 286 (J1331+305) as a flux calibrator and J1717-3948 as a phase calibrator (with a cycle time of 9 minutes on source, and 1 minute on the calibrator). 
\begin{figure}
\plotone{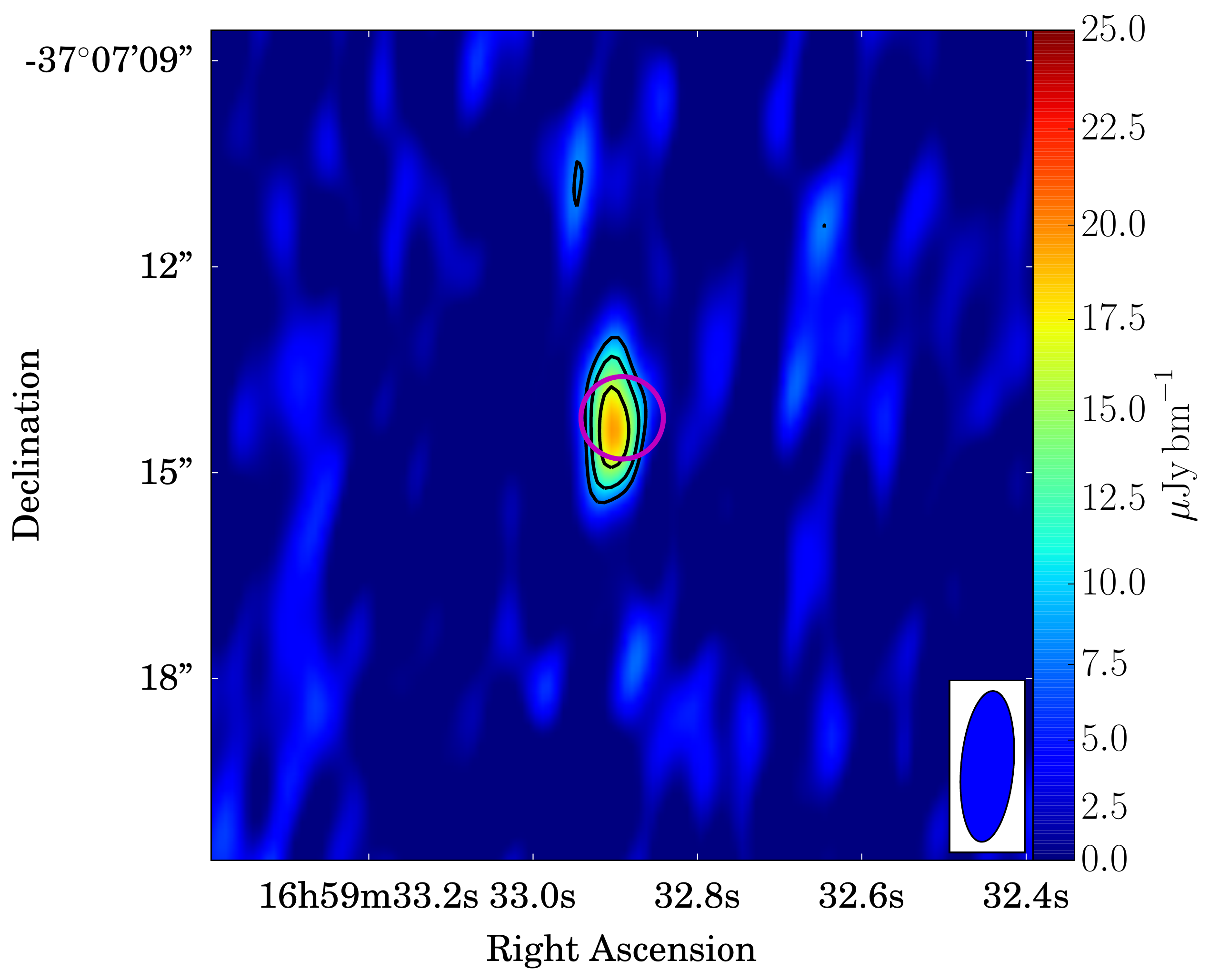}
\caption{\label{fig:vla} VLA radio image of \IGR{} taken at 10 GHz. We produced this image by stacking the data of both VLA epochs in the uv-plane. The source is clearly detected in the image, where contour levels are $2^{n/2}$ $\times$ the RMS noise of $2.8\mu{\rm Jy\,bm}^{-1}$, with $n=3,4,5$. The color bar represents the flux density in units of $\mu{\rm Jy\,bm}^{-1}$, and the blue ellipse is the VLA beam (the elongated beam shape is due to the low declination of \IGR{}). The \chandra{} (pink circle) X-ray error region is also shown, indicating that the VLA and \chandra{} localizations of the source are consistent.}
\end{figure}

We significantly detect a radio source at a position consistent with the \swift{} X-ray position reported in \citet{boz17b} {(see Figure~\ref{fig:vla})}. In the combined 4 GHz of bandwidth centered on 10 GHz, we measure flux densities of $17.7\pm4.4\,\mu {\rm Jy}$ and $24.4\pm4.3\,\mu {\rm Jy}$ on October 23 and 27, respectively.
To measure these flux densities we fit a point source in the image plane (with the \texttt{imfit} task).
The fluxes were too low to obtain a meaningful constraint on the radio spectral index using images of the two individual basebands.
The corresponding radio luminosities are shown in Table~\ref{table:obs_times}.

Additionally, we searched for intra-observation variability within both epochs of VLA data, on timescales as short as 30 min, which is the shortest timescale we can probe given the low source brightness. In both observations, the variance in the data points is consistent with the measurement uncertainties; we thus find no statistically significant evidence for flux variability on intra-observation timescales.

\subsection{Swift X-ray observations of \IGR{}}
\label{sec:xrt}
\IGR{} was observed with \swiftxrt{} twice following its detection with \integral{}.
These observations occurred on 2017 October 22 in photon counting mode (PC; which produces 2-dimensional images), and 2017 October 25 in windowed timing mode (WT; which collapses data to 1-dimension for fast readout). Observation times are displayed in Table~\ref{table:obs_times}.

We used \textsc{heasoft} v6.22 and \textsc{ftools}\footnote{\url{http://heasarc.gsfc.nasa.gov/ftools/}} \citep{black95} for all data reduction and analysis. All \swiftxrt{} observations were reprocessed via \texttt{xrtpipeline}, and \texttt{xselect} was used to manually extract source and background spectra. We used \texttt{xrtmkarf} to produce ancillary response files. 
Finally, we performed spectral analysis using \textsc{xspec} 12.9.1n \citep{ar96}. 

\renewcommand\tabcolsep{6pt}
\begin{deluxetable}{cccc}
\tablewidth{0pt}
\tablecaption{Best-fit X-ray spectral fitting parameters for \IGR{}\label{table:xspec}}
\tablehead{%
\colhead{Epoch} &
\colhead{$N_H$\tablenotemark{a}} &
\colhead{Photon} & \colhead{$F_{1-10{\rm keV}}$\tablenotemark{b}}\\
\colhead{} & \colhead{(${\times10^{22}{\rm \, cm}^{-2}}$)} &
\colhead{Index} &
\colhead{$({\times10^{-10}{\rm \, erg\,s^{-1} \,cm^{-2}}})$}
}
\startdata
1&$1.5\pm0.2$&$1.5\pm0.1$&$2.73\pm0.11$\\
2&$1.1\pm0.1$&$1.3\pm0.1$&$2.80\pm0.08$\\
\enddata
\tablenotetext{a}{Absorption column density.}
\tablenotetext{b}{Unabsorbed 1--10 keV flux.}
\end{deluxetable}
\renewcommand\tabcolsep{6pt}

\renewcommand\tabcolsep{2pt}
\begin{deluxetable}{ lcccc }
\tablecaption{Radio and X-ray luminosities of additional neutron star sources\label{table:other_srcs}}
\tabletypesize{\footnotesize}
\tablehead{
\colhead{Source}&
\colhead{$L_{5 {\rm GHz}}$\tablenotemark{a,b}}
&
\colhead{$L_{1\mbox{--}10 {\rm \, keV}}$}&
\colhead{D\tablenotemark{c}} &
\colhead{Ref.}
\\
\colhead{}&
\colhead{(${\rm \, erg\, s}^{-1}$)}&
\colhead{(${\rm \, erg\, s}^{-1}$)}&
\colhead{(kpc)}&
\colhead{}
}
\startdata
MAXI J0911$-$635&$<4.5\times10^{28}$&$2.5\times10^{36}$&10.4&[1]\\[0.1cm]
 SAX J1748.9$-$2021&$<4.5\times10^{27}$&$2.1\times10^{36}$&8.5&[2]\\[0.1cm]
 &$<5.1\times10^{28}$&$3.0\times10^{37}$&&[3]\\[0.1cm]
 Swift J175233.9-290952 &$<5.7\times10^{27}$&$1.4\times10^{34}$&8.0&[4]\\[0.1cm]
 4U 1543$-$624&$<7.2\times10^{27}$&$1.7\times10^{37}$&6.7&[5]\\[0.1cm]
 MAXI J0556$-$332&$\phm{<}5.3\times10^{28}$&$1.8\times10^{37}$&8.0&[6]\\[0.1cm]
 MXB 1730$-$335\tablenotemark{d}&$\phm{<}2.0\times10^{28}$&$6.1\times10^{35}$&8.6&[7]\\[0.1cm]
&$\phm{<}1.6\times10^{29}$&$4.0\times10^{37}$&&\\[0.1cm]
&$\phm{<}1.3\times10^{29}$&$4.1\times10^{37}$&&\\[0.1cm]
&$\phm{<}1.5\times10^{29}$&$7.2\times10^{37}$&&\\[0.1cm]
&$\phm{<}1.8\times10^{28}$&$3.3\times10^{36}$&&\\[0.1cm]
\enddata
\tablerefs{[1] \citet{tud16atel}; [2] \citet{teta17a}; [3] \citet{mj10atel}; [4] \citet{tet17atel}; [5] \citet{lud17atel}; [6] \citet{cor11atel}; [7] \citet{rut98atel}.}

\tablenotetext{a}{Upper limits are quoted at the $3\sigma$ level.}
\tablenotetext{b}{We calculate 5 GHz radio luminosities ($L_R=\nu L_\nu$) by assuming a flat spectral index to extrapolate to 5 GHz.}
\tablenotetext{c}{Distance value used to calculate luminosity.}
\tablenotetext{d}{Note that we use a model with $N_{\rm H}=1.7\times10^{22}\,{\rm cm}^{-2}$ \citep{marsha01}, and a photon index of 1.5, to convert from RXTE count rates to flux for this source.}

\end{deluxetable}
\renewcommand\tabcolsep{6pt}

{The PC mode observation was piled up. Therefore, we followed the recommended procedure for handling pile up\footnote{\url{http://www.swift.ac.uk/analysis/xrt/pileup.php}}, and extracted
a source spectrum from an annulus, excluding the piled up region (estimated to be $\sim10$ arcsec), out to $80$ arcsec. Following the recommended procedure for \swiftxrt{} data analysis, we used events from grades 0-12, and performed spectral analysis in the 0.3--10 keV band.

For the WT mode observation we used circular regions with a radius of $\sim47$ arcsec (20 pixels) for both source and background. To minimize the effects of WT spectral residuals\footnote{For more details on these effects see the \swiftxrt{} calibration digest; \url{http://www.swift.ac.uk/analysis/xrt/digest\_cal.php\#abs}}, we only extracted a spectrum from grade 0 events and performed spectral analysis in the 0.5-10 keV band, as these residuals become prominent around and below 0.5 keV.}

 \begin{figure*}
\centering
 {\plotone{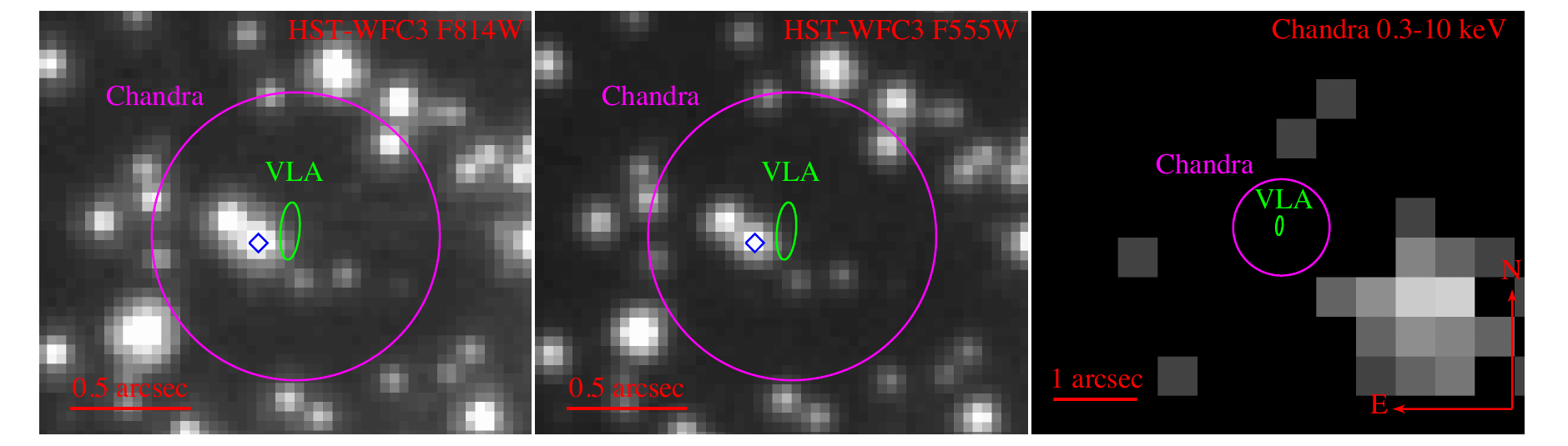}}
\caption{\label{fig:hstch} Archival \hst{} and \chandra{} images of the field surrounding \IGR{}. The left panel displays the \hst{} F814W image, the middle panel displays the \hst{} F555W image, and the right panel displays the \chandra{} (0.3--10 keV band) image. The \chandra{} (magenta circles), and VLA (green ellipses) error regions are indicated in all panels, and the blue diamonds indicate the possible counterpart discussed in \S\ref{sec:counterpart}. Note the different scales in the optical ({\it left} and {\it middle}) and X-ray ({\it right}) images. We do not detect an optical or quiescent X-ray counterpart to \IGR{} in these archival data.}
\end{figure*}

We extracted the X-ray spectrum from each XRT observation separately to perform spectral fitting. Both of the XRT spectra are well fit with an absorbed power-law ({\sc tbabs}*{\sc pegpwrlw} in \textsc{xspec}), where we assume photo-electric cross sections from \citet{vern96} and abundances from \citet{wilms00}. {We chose to use the {\sc tbabs} ISM absorption model, as this model implements more recent estimates for the elemental abundance of the ISM, when compared with older models (e.g., {\sc phabs} or {\sc wabs}; \citealt{bahramian2015,fo16}).} The best fit spectral fitting parameters for both epochs are shown in Table~\ref{table:xspec}, and the corresponding X-ray luminosities are shown in Table~\ref{table:obs_times}. {We report $1\sigma$ confidence intervals on all fitted parameters. To calculate these $1\sigma$ confidence intervals, 
we first binned each spectrum to a minimum of 50 counts per bin using the \texttt{grppha} task, and utilized $\chi^2$ statistics for spectral analysis. Following this analysis, the parameter uncertainties were estimated using the \texttt{error} task in {\sc xspec} \footnote{For details on how this task works see \url{https://heasarc.gsfc.nasa.gov/xanadu/xspec/manual/node80.html}}}.

\subsection{Archival Optical and X-ray observations}
We obtained archival X-ray and optical data of the field surrounding \IGR{} ({see Figure~\ref{fig:hstch}}), taken with the \chandra{} X-ray Observatory and the {\em Hubble Space Telescope} (\hst).
\chandra{} data were taken on 2008 January 26 (13:09:51 UTC start time, 9.4 ks exposure time, {Obs ID: 8951}) with the ACIS-S instrument. We reprocessed the data using \textsc{ciao} v4.9 \citep{ciao}.
\hst{} images were obtained from the Hubble Legacy Archive (\url{http://hla.stsci.edu/}) in the F555W and F814W filters. These \hst{} observations were taken with the WFC3/UVIS1 detector on 2009 August 02 (19:47:12/20:43:35 UTC start time, 1.1/0.3 ks exposure time, in the F555W/F814W filters).
The absolute astrometry of the \hst{} images was corrected by matching sources in the field to the GAIA catalog \citep{gaia}. We estimate that after matching to the GAIA catalog, the
uncertainty in the absolute position registration of the \hst{} images is $< 0\farcs02$.

\subsection{Additional neutron star sources}
We supplement\footnote{The additional neutron star data reported in this section have only been reported in Astronomer's Telegrams, and not previously published in refereed journals.} these new radio/X-ray data on \IGR{} with our team's recent radio/X-ray measurements of a number of other neutron star sources for further comparison, MAXI J0911-635 \citep{tud16atel}, SAX J1748.9-2021 \citep{mj10atel,teta17a}, Swift J175233.9-290952 \citep{tet17atel}, and 4U 1543-624 \citep{lud17atel}. Additionally, we also include older detections of MAXI J0556-332 \citep{cor11atel} and MXB 1730-335 \citep{rut98atel} in this work.
Table~\ref{table:other_srcs} displays a summary of the radio and X-ray luminosity measurements for these sources.

\section{Results}

\subsection{Radio source position}
\label{sec:rad_pos}
Stacking both epochs of our VLA data in the {\it uv}-plane (see Figure~\ref{fig:vla}) refines the radio position of \IGR{} to be the following (J2000),
\begin{eqnarray}\nonumber 
{\rm RA:}& \phantom{-}16^{\rm h}59^{\rm m}32\fs90230\pm 0\fs00092 \pm 0\fs005 \\ \nonumber
{\rm DEC:}& -37^{\circ}07'14\farcs278\phn\phn\pm 0\farcs088\phn\phn \pm 0\farcs22\phn, \nonumber
\end{eqnarray}
where the quoted errors represent the statistical error from fitting in the image plane and the nominal systematic uncertainties of 10 percent of the beam size, respectively. The elongated beam shape arises from the low declination of the source.
This radio source position is
consistent (within $0\farcs04$) with the best X-ray position of the source during outburst from \chandra{} \citep{chak17}.

\subsection{Search for the quiescent X-ray and optical counterparts}
\label{sec:counterpart}
We examined archival \chandra{} 
and \hst{} 
observations in search of the quiescent X-ray and optical counterparts to \IGR{}.

In the archival \chandra{} data, there is an X-ray source $\sim2.2$ arcsec to the south-west of the VLA radio position (see Figure~\ref{fig:hstch} \textit{right panel}). However, this \chandra{} source is unlikely to be the quiescent counterpart of IGR J16597-3704 given the typical \chandra{} absolute astrometric accuracy of $0.5$ arcsec. 
To further confirm the \chandra{} absolute astrometry is accurate (i.e., as good as $0.5$ arcsec or better), we compare the positions of known X-ray sources in the cluster to their radio and optical counterparts.
In particular, there is a radio continuum source $\sim 2.5$ arcmin from the center of the cluster that has a \chandra{} X-ray counterpart, and the positions of these match to within $0.5$ arcsec. Additionally, there are three X-ray sources in the outer regions of the cluster, where the optical source density is lower, that all clearly match bright stars present in the \emph{Gaia} catalog. The individual offsets between the \emph{Gaia} and \chandra{} positions vary from 0.6--$1.0$ arcsec, well within the uncertainties of the individual X-ray positions, and there is no evidence of a significant net astrometric shift.
Together these arguments strongly suggest that the bright X-ray source in question is not associated with IGR J16597-3704, and that the quiescent counterpart of this transient is undetected in existing X-ray data.

We assert a non-detection in this \chandra{} observation and estimate a 95\% upper limit on the count rate of $2.9\times10^{-4}\, {\rm cts\,s}^{-1}$. Assuming a distance of $D=9.1$ kpc \citep{val07a}, hydrogen column density of $\sim1.1\times10^{22}\,{\rm cm}^{-2}$, and power law spectrum (with a canonical photon index of 1.5), this translates to upper limits of $4.9\times10^{-15}$~erg~s$^{-1}$~cm$^{-2}$
on the absorbed X-ray flux in the 0.5-10 keV band, and $L_X < 6.4\times10^{31}$ erg s$^{-1}$ for the luminosity, of the quiescent counterpart.
Alternatively, assuming an neutron star atmosphere model ({\sc nsatmos} in {\sc xspec}) with canonical values of $1.4 M_\odot$ and a radius of $10$ km, this translates to upper limits of $1.1\times10^{-15}$~erg~s$^{-1}$~cm$^{-2}$
on the absorbed X-ray flux in the 0.5-10 keV band, and $L_X < 9.9\times10^{30}$ erg s$^{-1}$ for the luminosity.
In this model, the upper limit on the absorbed flux corresponds to a neutron star temperature of  $< 75\,{\rm eV}$.
{Neutron star temperatures have been measured in other systems to extend across a range of values, from $<50\, {\rm eV}$ for the coolest neutron stars (e.g., EXO 1745$-$248, SAX J1808.4$-$3658, 1H 1905$+$000; \citealt{jonk07,hein09,deg12}), up to $\sim150\,{\rm eV}$ for the hottest neutron stars (e.g., XTE J1701-462; \citealt{wij17}). Our new temperature measurement indicates that the neutron star in \IGR{} is not an overly hot neutron star, but rather is consistent with an average or lower temperature neutron star, when compared with the current measured population.}

 \begin{figure*}
\centering
 {\includegraphics[width=2\columnwidth]{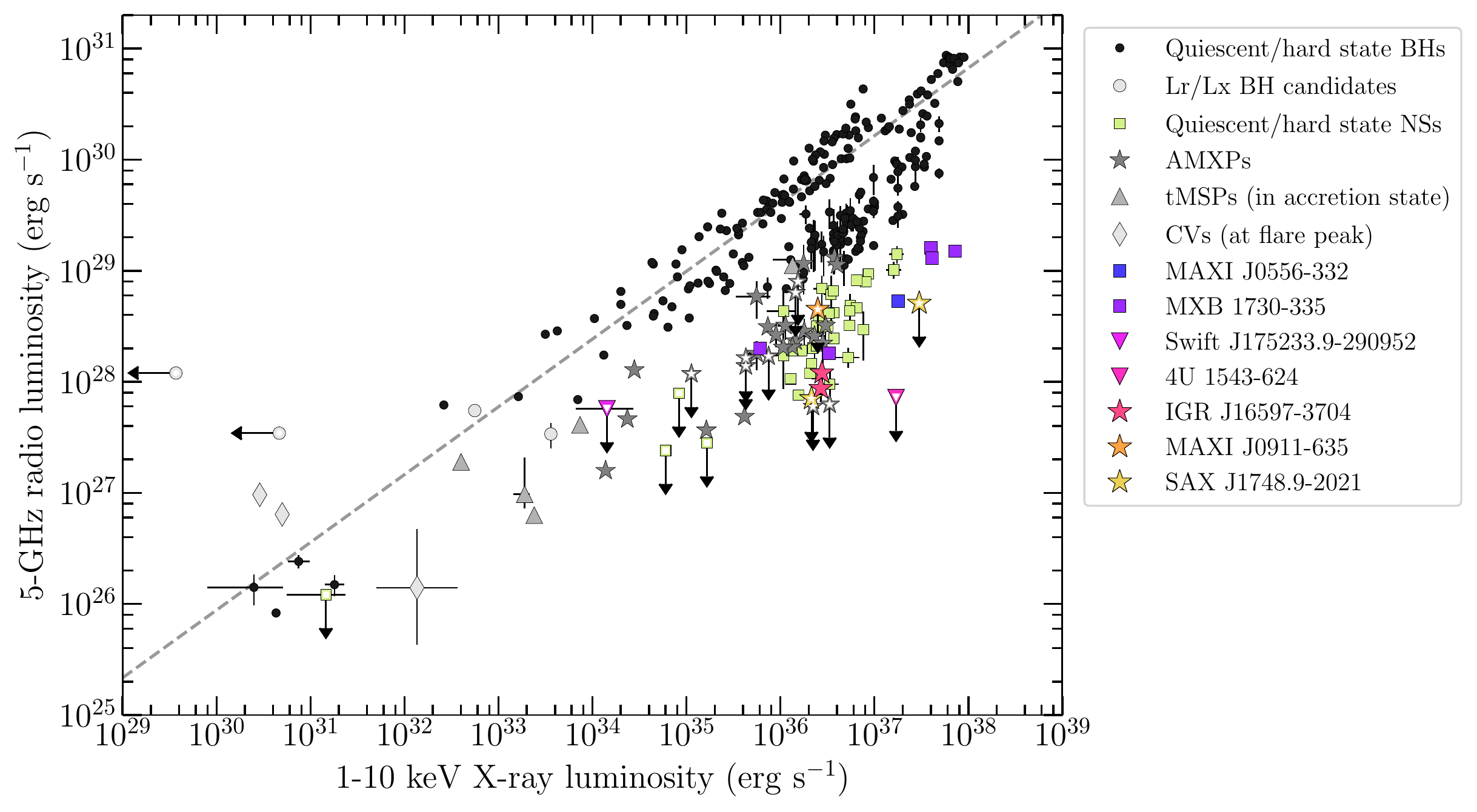}}
\caption{\label{fig:lrlx}The radio -- X-ray correlation for different types of accreting stellar-mass compact objects: black holes; different classes of neutron stars (non-pulsating NSXBs, AMXPs, and tMSPs); and out-bursting cataclysmic variables. This plot is adapted from \cite{bahr17}, with additional measurements from more recent publications, as well as those displayed in Table~\ref{table:other_srcs}; \citealt{rut98atel,gallo06,mj10atel,cor11atel,rus15,marsh16,tetarenkoa2016,rush16,tud16atel,ribo17,plot17,tud17,teta17a,gus17,tet17atel,lud17atel,bog17,din17}. The best-fit relation for black holes (${\beta=0.61}$, grey dashed; \citealt{gallo14}) is also shown.
Our measurements of \IGR{} and the measurements of other NSXBs reported in Table~\ref{table:other_srcs} are displayed with colored symbols (where non-pulsating NSXBs, AMXPs, and unclassified sources are indicated by the square, star, and inverted triangle shapes, respectively).
\IGR{} (pink star shapes)
is one of the more radio quiet systems in the NSXB population. }
\end{figure*}

In the \hst{} data, there is no clear optical source within the VLA error circle (see Figure~\ref{fig:hstch} \textit{left} \& \textit{middle}). However, we identify a bright optical source 0\farcs13 to the SE of our VLA radio position, which lies outside the $1\sigma$ VLA confidence interval. Optical photometry indicates this \hst{} source was at ${\rm m}_{\rm F555W/F814W}\sim {22.4/20.3}$ (AB magnitudes) on 2009 August 02, consistent with a typical giant star within the NGC 6256 cluster. Given that \IGR{} has been recently identified \citep{sanna17} to be an ultra-compact system (which typically have compact white dwarf companions), this \hst{} source is unlikely to be the optical counterpart. Therefore, the optical counterpart is probably too faint to be detected in the existing \hst{} data. We estimate $3\sigma$ upper limits from the archival \hst{} images of ${\rm m}_{\rm F555W/F814W}< {26.0/23.4}$.

\subsection{Radio -- X-ray correlation}
\label{sec:lrlx}
To explore the nature of \IGR{}, we place our observations on the radio -- X-ray plane, using the 5 GHz radio luminosity and the $1.0-10$ keV X-ray luminosity (where that frequency and band are chosen to match measurements from the literature; see Figure~\ref{fig:lrlx} \& Table~\ref{table:obs_times}).

The location of \IGR{} on the radio -- X-ray plane lies at least an order of magnitude below most BHXBs, and instead is more consistent with neutron star systems (both non-pulsating NSXBs and AMXPs/tMSPs; see Figure~\ref{fig:lrlx}). The recent detection of X-ray pulsations from this source with NuSTAR \citep{sanna17}, confirms that IGR J16597-3704 is a new AMXP source. All of the neutron star sources presented in this work display a significant range in radio luminosity. In particular, \IGR{}, along with SAX J1748.9-2021, Swift J175233.9-290952, and 4U 1543-624, display radio luminosities that are at the low end of the sampled NSXB population, while MAXI J0911-635, MAXI J0556-332, MXB 1730-335 display radio luminosities at the mid to high end
of the sampled NSXB population.

\section{Discussion}
\label{sec:discuss}
In this paper, we have reported on the discovery of the radio counterpart to the new AMXP, \IGR{}, located in the NGC 6256 globular cluster.  We do not conclusively identify an optical or quiescent X-ray counterpart to \IGR{} in archival \hst{} and \chandra{} data; our $3\sigma$ upper limits are ${\rm m}_{\rm F555W/F814W}< {26.0/23.4}$ and $L_X < 6.1\times10^{31}$ erg s$^{-1}$.

Our recent radio observations indicate that IGR J16597$-$370 is one of the more radio faint systems in the NSXB population. 
For example, \IGR{} displays a similar radio luminosity to IGR~J17511$-$3057 (AMXP; \citealt{tud17}), SAX~J1748.9$-$2021 (AMXP; \citealt{teta17a}), and EXO~1745$-$248 (non-pulsating NSXB; \citealt{tetarenkoa2016}). Examining our updated radio -- X-ray plane figure (Figure~\ref{fig:lrlx}), it is clear that both non-pulsating NSXBs and AMXPs can display a range of radio luminosities at a similar X-ray luminosity (where it is unclear whether the radio-brighter or radio-fainter systems form the dominant population).
Here we postulate on the mechanisms driving the radio luminosity in \IGR{} (and potentially other AMXPs), by exploring the relationships between radio luminosity and spectral state, spin, magnetic field, orbital period, accretion regime, and evolutionary state.

The jets from some NSXBs (like BHXBs) have been observed to be quenched by over an order of magnitude (or faded below current detection limits; e.g., \citealt{mig03,gus17}) during softer accretion states.
As such, we may naively expect radio jets to be fainter in these states when compared to their harder accretion states.
Since, the X-ray spectral properties reported in \S\ref{sec:xrt} suggest that \IGR{} was in a canonical hard state during our observations, it is unlikely that jet quenching in the soft accretion state explains the low radio luminosity of \IGR{}. By extension, many other radio quieter systems in the sampled population, such as EXO~1745$-$248 \citep{tetarenkoa2016}, have also been observed firmly in the hard accretion state.

\citet{sanna17} have  shown that \IGR{} displays a longer spin period ($9.5$ ms), when compared to the average values for AMXPs \citep{patr12,muk15,patr17b,patr17}. 
{The spin of a neutron star has long been suggested to potentially affect the radio luminosity of NSXBs \citep{migl11b}.
Depending on how the magnetic field (anchored to the NS magnetic poles) interacts with the accretion disc, AMXP jets could be directly powered by the extraction of energy from the spin of the neutron star, or the jet may be driven by the rotation power of the accretion disc \citep{migl11b,migl11a}. In both cases, we expect the neutron star spin period to correlate with jet power. For example, a longer spin period could be linked to a lower radio luminosity in AMXPs, analogous to the spin dependence of jet power for black holes ($L_{\rm jet}\propto a^2$, where $a$ is the black hole spin parameter) predicted by \cite{bz}.} 
Past studies \citep{migl11b} have found hints of a possible positive correlation\footnote{Although, we note that the jet power was not measured directly in this work, with the normalization of the sources on the radio X-ray plane (assuming a disc-jet coupling index of 1.4) being used as a proxy for jet power. Thus this correlation may break down for different disc-jet coupling indices.} between spin frequency of the neutron star and jet power in AMXPs, and our measurements of \IGR{} are compatible with this scaling.
Therefore, it seems plausible that spin period may play a role in governing the radio luminosity levels in \IGR{} (and potentially other radio-quiet AMXPs).

Similar to the spin period, \IGR{} may also display a higher magnetic field ($9.2\times10^8<B<5.2\times10^{10}$ G) when compared to the average values for AMXPs. The role of a high magnetic field in jet production is still an open question.
Past works have suggested that high magnetic fields ($\gtrapprox10^{11}$ G; \citealt{fenhen00,migl11a}) may inhibit jet formation, but the recent work by \citealt{vde17a,vde17b} (described below) provides a counterpoint, thus we explore both possibilities here.
Using the condition that the gas pressure must dominate over the magnetic pressure, \cite{mas08} derive a condition for jet formation based on the magnetic field strength ($B_*$) and accretion rate ($\dot{M}$), such that,
\begin{equation}
\frac{R_A}{R_*}=0.87\left(\frac{B_*}{10^8 \,{\rm G}}\right)^{4/7}\left(\frac{\dot{M}}{10^{-8} \,M_\odot{\rm yr}^{-1}}\right)^{-2/7}
\end{equation}
where $R_A$ is the Alfv\'en radius, and $\frac{R_A}{R_*}\approx1$ indicates the portion of the parameter space where jet formation is likely to not be suppressed by the neutron star magnetic field. Substituting in estimates of $9.2\times10^8<B_*<5.2\times10^{10}$ G \citep{sanna17} and $\dot{M}=5\times10^{-12}\,M_\odot{\rm yr}^{-1}$ (estimated from the $\dot{M}$/$P_{\rm orb}$ relationship reported in \citealt{vanh12}\footnote{While \cite{sanna17} report an $\dot{M}=5.5\times10^{-10}\,M_\odot\,{\rm yr}^{-1}$, this estimate only represents the peak $\dot{M}$ in the outburst.}) 
for \IGR{},  indicates that \IGR{} may be in a regime ($\frac{R_A}{R_*}>1$) where the magnetic field could potentially be inhibiting jet formation (and in turn lead to the lower radio luminosity observed). 

Contrary to this hypothesis, radio emission (consistent with a synchrotron jet) has been recently detected in the high magnetic field neutron star systems GX 1$+$4 \citep{vde17a} and Her X-1 \citep{vde17b}. Moreover, another AMXP, IGR~J17511$-$43057, displays a magnetic field strength similar to the average AMXP population, but lower than average radio luminosity \citep{tud17}.
Furthermore, if high magnetic fields are linked to lower radio luminosities in neutron stars, we may expect that the AMXP population in general would display
lower radio luminosities compared to the population of non-pulsating NSXBs (which presumably display lower magnetic fields than pulsating systems). This is clearly 
not the case, as for example, the non-pulsating NSXB, EXO 1745$-$248, displays a radio luminosity similar to \IGR{} (also see Figure ~\ref{fig:lrlx}).
Therefore, there does not appear to be a clear relationship between magnetic field strength of the neutron star and radio luminosity in the current sampled population, suggesting that the high magnetic field in \IGR{} does not strongly influence the radio luminosities we observe.

\renewcommand\tabcolsep{2pt}
\begin{deluxetable}{ lccccc }
\tabletypesize{\footnotesize}
\tablecaption{Properties of ultra-compact neutron star binaries\label{table:ucxb_tab}}
\tablehead{
\colhead{Source}&
\colhead{$L_{5 {\rm GHz}}$\tablenotemark{a,b}}&
\colhead{$L_{1\mbox{--}10 {\rm \, keV}}$\tablenotemark{a}}&
\colhead{D\tablenotemark{c,d}} &
\colhead{$P_{\rm orb}$\tablenotemark{d}} &
\colhead{Ref.}
\\
\colhead{}&
\colhead{(${\rm \, erg\, s}^{-1}$)}&
\colhead{(${\rm \, erg\, s}^{-1}$)}&
\colhead{(kpc)}&
\colhead{(min)}&
\colhead{}
}
\startdata
  4U1728-34&$6.83\times10^{28}$&$5.2\times10^{36}$&5.2&10.8?\tablenotemark{$\ddagger$}&[1]\\[0.1cm]
  4U~1820-303&$8.78\times10^{28}$&$9.7\times10^{36}$&7.9&11&[1]\\[0.1cm]
  4U~0513-40&$<5.50\times10^{28}$&${2.9\times10^{36}}$\tablenotemark{*}&12.1&17&[2]\\
  2S0918-549&$<5.21\times10^{28}$&${9.5\times10^{35}}$\tablenotemark{*}&5.4&17.4&[3]\\
  4U~1543-624&$<7.20\times10^{27}$&$1.7\times10^{37}$&6.7&18.2&[4]\\[0.1cm]
  4U~1850-087&$4.60\times10^{28}$&${1.2\times10^{36}}$\tablenotemark{*}&6.9&20.6&[5]\\
  M15 X-2&$3.67\times10^{28}$&$2.3\times10^{37}$&10.4&22.6&[6]\\[0.1cm]
  4U~1916-053&$<1.80\times10^{29}$&${2.7\times10^{36}}$\tablenotemark{*}&9.3&50&[7]\\
  4U~0614+091&$1.72\times10^{28}$&$3.2\times10^{36}$&3.2&51?\tablenotemark{$\ddagger$}&[8]\\
  XTE~J1751-305&$<1.14\times10^{28}$&${<2.3\times10^{32}}$\tablenotemark{$\dagger$}&8.0&42&[9]\\
  XTE~J0929-314&$1.37\times10^{29}$&$4.7\times10^{36}$&8.0&43.6&[10]\\
\enddata
\tablerefs{[1] \citealt{dt17}; [2] \citealt{mach90}; [3] \citealt{zw93}; [4] \citealt{lud17atel}; [5] \citealt{leh90}; [6] \citealt{siva11}; [7] \citealt{grin86}; [8] \citealt{miga10}; [9] \citealt{ia10}; [10] \citealt{rup02}}
\tablenotetext{a}{Upper limits are quoted at the $3\sigma$ level.}
\tablenotetext{b}{We calculate 5 GHz radio luminosities ($L_R=\nu L_\nu$) by assuming a flat spectral index to extrapolate to 5 GHz.}
\tablenotetext{c}{Distance value used to calculate luminosity.}
\tablenotetext{d}{All distance and orbital period measurements are taken from \citet{cart13}.}
\tablenotetext{*}{These systems did not have X-ray measurements reported with their radio measurements; we place limits on the X-ray luminosity by using the luminosity functions reported in \citet{cart13}.}
\tablenotetext{\dagger}{XTE~J1751$-$305 did not have an X-ray measurement reported with its radio measurement; we use the upper limit on the quiescent X-ray luminosity reported in \citet{w05}.}
\tablenotetext{\ddagger}{\citet{cart13} classify these estimates of orbital period as more uncertain, as they are supported by only weak evidence.}

\end{deluxetable}
\renewcommand\tabcolsep{6pt}

Although, if different jet production mechanisms are at work in different classes of neutron star systems, we may expect a much more complicated (beyond a simple scaling) relationship between the magnetic field strength of the neutron star and the radio luminosity. For instance, the jet production mechanism in neutron star systems could be highly dependent on how dynamically important the magnetic field of the neutron star is in each system (i.e, how significant a role the stellar magnetic fields play in the accretion process). In this case, jets launched from non-pulsating NSXBs, with dynamically unimportant magnetic fields, may be powered by the accretion disc, similar to BHXBs, while the dynamically important magnetic fields in tMSPs/AMXPs could disrupt this physical connection between the jet and the disc. Therefore, in some systems we may not be observing an accretion powered jet, but rather another mechanism, such as the propeller effect\footnote{In the case of the propeller effect, the radio emission may originate in a broader outflow, as opposed to a well-collimated jet.} \citep{rom09,par17}, which may be powering the jet (e.g., the propeller effect is thought to be the origin of the anti-correlation between radio and X-ray luminosity observed in PSR J1023+0038; \citealt{bog17}).

\begin{figure}
\centering
 {\includegraphics[width=1\columnwidth]{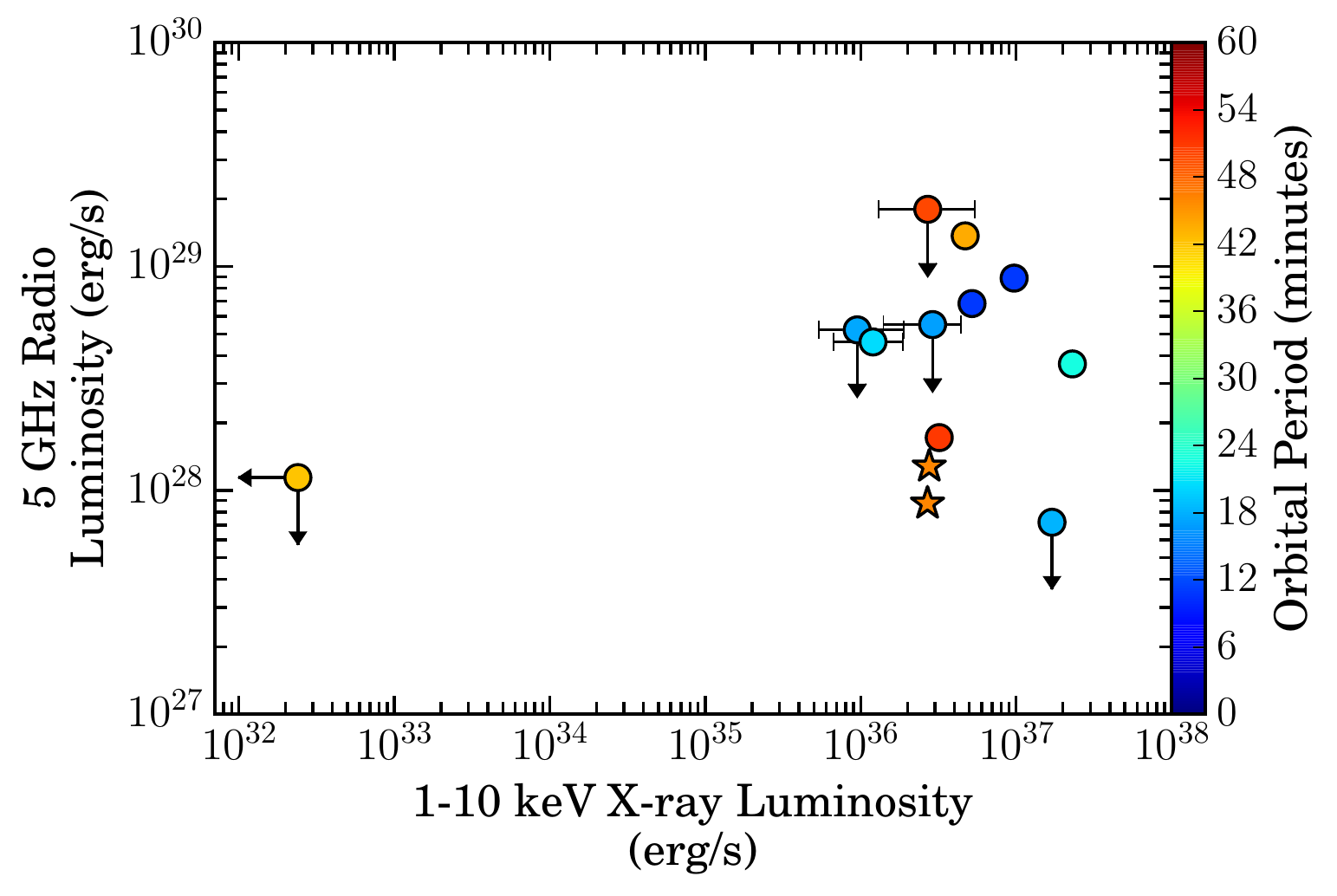}}
\caption{\label{fig:ucxb_fig} Radio -- X-ray correlation for 
ultra-compact neutron star binaries. The data points are color coded by orbital period.
Our \IGR{} measurements are indicated by the star symbols. The data points with horizontal error bars represent those systems for which the X-ray luminosity measurement were estimated from luminosity functions or quiescent X-ray limits (see Table~\ref{table:ucxb_tab} for details). We observe no clear correlation between the position of these sources on the radio -- X-ray plane and orbital period.}
\end{figure}

As \IGR{} is an ultra-compact binary (with an orbital period $<80$ min), we opt to briefly investigate a possible link between radio luminosity and orbital period, by compiling a list of all the ultra-compact neutron star binaries with radio frequency measurements (see Table~\ref{table:ucxb_tab} and Figure~\ref{fig:ucxb_fig}). We find that in these ultra-compact binaries, the orbital period does not appear to be correlated with the position of the system in the radio -- X-ray plane. Further, we also find no evidence of a direct correlation between orbital period and radio luminosity (Spearman rank correlation coefficient of $-0.22$, and p-value of 0.46).
This suggests that orbital period may not play a key role in governing the radio luminosity in \IGR{} or other systems.

Lastly, the radio luminosity in AMXPs may be closely tied to the evolutionary state or the accretion regime (i.e., X-ray spectral state and mass accretion rate, as suggested by \citealt{migl11b}) 
of the system.
In \IGR{}, \citet{sanna17} estimate that very little mass has been accreted so far in the system, and suggest that this indicates it is in an early stage of its evolution (i.e., it is a partially-recycled pulsar). However, the likelihood of catching this system in such a short-lived early evolutionary state is quite low. Since \IGR{} is an ultra-compact binary (with a low donor star mass; \citealt{sanna17}), it is much more consistent with a system in the later stages of its evolution. This uncertainty makes it difficult to postulate whether the evolutionary state of \IGR{} (and other AMXPs) influences their observed radio luminosities, without further study.

A larger sample of radio luminosity constraints from AMXPs is needed to definitively determine whether the spectral state, spin period, magnetic field strength, orbital period, or accretion regime/evolutionary state are linked to jet behaviour.

Overall, these results highlight the need for more radio and X-ray measurements of all classes of NSXBs to place improved constraints on the mechanisms that govern radio luminosity, jet production and jet evolution in NSXBs. The low X-ray luminosity regime ($L_X<10^{36}\, {\rm erg\, s}^{-1}$) is particularly vital, as this regime remains under-sampled for the different classes of NSXBs.
Finally, despite the lack of a clear correlation for neutron star systems, \IGR{} is a clear example that the radio -- X-ray plane can still be a reliable diagnostic to identify the nature of the accretor in these binary systems.



\acknowledgments
We thank the National Radio Astronomy Observatory and Swift staff for rapidly scheduling the observations reported in this paper. The authors thank Lennart van Haaften for a helpful discussion on binary evolution models and Vlad Tudor for his help compiling radio and X-ray measurements of additional neutron star systems for this work.
AJT is supported by an Natural Sciences and Engineering Research Council of Canada (NSERC) Post-Graduate Doctoral Scholarship (PGSD2-490318-2016). AJT, GRS, COH, and AWS are supported by NSERC Discovery Grants. JCAMJ is the recipient of an Australian Research Council Future Fellowship (FT140101082). RMP acknowledges support from Curtin University through the Peter Curran Memorial Fellowship. ND is supported by a Vidi grant from Netherlands Organisation for Scientific Research (NWO). TDR acknowledges support from the NWO Veni Fellowship. JS acknowledges support from the Packard Foundation. JAK acknowledges support from the NASA contract NAS5-00136. The National Radio Astronomy Observatory is a facility of the National Science Foundation operated under cooperative agreement by Associated Universities, Inc. This work has made use of data from the European Space Agency (ESA)
mission {\it Gaia} (\url{https://www.cosmos.esa.int/gaia}), processed by
the {\it Gaia} Data Processing and Analysis Consortium (DPAC,
\url{https://www.cosmos.esa.int/web/gaia/dpac/consortium}). Funding
for the DPAC has been provided by national institutions, in particular
the institutions participating in the {\it Gaia} Multilateral Agreement.
This research has made use of the following data and software packages: Swift XRT Data Analysis Software (XRTDAS) developed under the responsibility of the ASI Science Data Center (ASDC), Italy. We acknowledge extensive use of the ADS and arXiv.



\facilities{VLA, Swift (XRT), HST (WFC3/UVIS1), Chandra (ACIS-S)}
\software{\textsc{casa}, \citep[v5.1.1;][]{mcmullin2007}, 
\textsc{heasoft} \citep[v6.2.2;][]{black95}, \textsc{ftools} \citep{black95}, 
\textsc{xspec} \citep[12.9.1n;][]{ar96}, \textsc{ciao} \citep[v4.9;][]{ciao}}

\bibliography{ABrefList}
\bibliographystyle{aasjournal}


\end{document}